\documentclass[runningheads,envcountsame]{llncs}
\usepackage[T1]{fontenc}
\usepackage{amssymb,latexsym,amsmath}
\usepackage{cite}
\usepackage{colonequals}
\usepackage{paralist}

\usepackage{hyperref}

\spnewtheorem*{principle}{Knowledge of Preconditions Principle}{\bfseries}{\itshape}

\renewcommand{\phi}{\varphi}
\newcommand{\Prop}{\mathsf{Prop}}
\newcommand{\ce}{\colonequals}
\newcommand{\cce}{\coloncolonequals}
\newcommand{\R}{R}
\newcommand{\RR}{\mathcal{R}}
\newcommand{\KK}{\mathcal{K}}
\newcommand{\HH}{\mathcal{H}}
\newcommand{\correct}[1]{\mathit{correct}_{#1}}
\newcommand{\kni}{\mathit{knight}}
\newcommand{\kna}{\mathit{knave}}
\newcommand{\Byzf}{\mathit{Byz}_{\!f}}
\newcommand{\creed}[3]{\mathbb{C}_{#1}^{#2 \setminus #3}}

\begin{document}

\title{Communication Modalities}
\author{Roman Kuznets\thanks{This research was funded by  the Austrian Science Fund (FWF) project ByzDEL (P33600).}\orcidID{0000-0001-5894-8724}}
\authorrunning{R. Kuznets}
\institute{TU Wien, Vienna, Austria\\
\email{roman@logic.at}
}
\maketitle              
\begin{abstract}
Epistemic analysis of distributed systems is one of the biggest successes among applications of logic in computer science. The reason for that is that agents' actions are necessarily guided by their knowledge. Thus, epistemic modal logic, with its knowledge and belief modalities (and group versions thereof), has played a vital role in establishing both impossibility results and necessary conditions for solvable distributed tasks. In distributed systems, knowledge is largely attained via communication. It has been standard in both distributed systems and dynamic epistemic logic to treat incoming messages as trustworthy, thus, creating difficulties in the epistemic analysis of byzantine distributed systems where faulty agents may lie. In this paper, we argue that handling such communication scenarios calls for additional modalities representing the informational content of messages that should not be taken at face value. We present two such modalities: \emph{hope} for the case of fully byzantine agents and  \emph{creed} for non-uniform communication protocols in general.

\end{abstract}
\section{Introduction}
In their paper \emph{On the Unusual Effectiveness of Logic in Computer Science}~\cite{HalpernHIKVV01}, Halpern~et~al. list ``the deployment of epistemic logic to reason about knowledge in multi-agent systems'' as one of the ``areas of computer science on which logic has had a definite and lasting impact'' and support their thesis with multiple examples. This should come as no surprise since knowledge is inextricably linked with action. Moses~\cite{Moses16} recently formalized this in the form of the \emph{Knowledge of Preconditions Principle}~(KoP):
\begin{principle}
If $\phi$~is a necessary condition for agent~$i$ performing action~$\alpha$, then $K_i \phi$~(i.e., agent~$i$'s knowledge of~$\phi$) is also a necessary condition for agent~$i$ performing action~$\alpha$.
\end{principle}

The logic commonly used  to reason  about knowledge is \emph{epistemic modal logic}~\cite{Hintikka62}, with its semantics of \emph{Kripke models} consisting of possible worlds. According to this semantics, knowledge is limited by how well  agents are able to  pinpoint the real world among the many worlds they consider possible: the fewer possibilities considered, the more the agent knows. 

This semantics is adapted to the specific needs of distributed systems in the \emph{runs and systems framework}~\cite{FaginHMV95}, with global states of the distributed system playing the role of possible worlds and with each agent's local states determining what this agent considers possible. As a result, all global states where the agent has the same local state are indistinguishable for the agent, resulting in rather strong properties of knowledge, including factivity (whatever is known must be true), as well as positive and negative introspection (the agent knows what it knows and what it does not know). In modal terms, this corresponds to (the multimodal version of) logic~{\sf S5}.

While the form of messages in distributed systems is not restricted, it is assumed that agents interpret each message they receive based on the common a~priori knowledge of the \emph{joint protocol}, i.e., of the protocols of each agent. This assumption of common a~priori knowledge of the protocols is akin to the common knowledge of the model pointed out by Artemov as an assumption in epistemic modal logic in general~\cite{Artemov20}. In particular,  it completely abstracts away any difficulties of interpreting messages correctly.

The other side of this coin, however, is that it is not clear how to interpret messages that are not sent according to a pre-determined protocol. A prominent example of this can be found in \emph{byzantine distributed systems}, i.e.,~distributed systems with \emph{fully byzantine} agents~\cite{LamportSP82}, i.e.,~agents that can arbitrarily deviate from their protocols. In particular, a fully byzantine agent can send any message at any time, independent of both its local state and its (correct) protocol. Thus, it should come as no surprise that, until recently, the standard epistemic analysis of distributed systems did not extend to byzantine distributed systems.\footnote{There were papers  about knowledge in ``byzantine'' or ``fault-tolerant'' distributed systems, such as \cite{MosesT88,DworkM90,HalpernMW01}, but the types of failures there is restricted to crashes and/or omissions, meaning that all messages are still sent according to the pre-determined protocol.}

Providing the epistemic analysis of distributed systems with fully byzantine agents was the goal of the project ``ByzDEL:~Reasoning about Knowledge in Byzantine Distributed Systems'' funded by the Austrian Science Fund~(FWF) that Roman Kuznets  led in TU~Wien with Ulrich Schmid as the~co-PI. In this paper, we  report (some of) the findings of our team. 

The difficulty of interpreting a message in byzantine systems is that the recipient must consider the possibility of the sender being  faulty. With faulty agents able to lie, there seems to be no increase in the recipient's knowledge. And yet distributed protocols work for byzantine systems. One of the basic observations is that, under the common assumption of having at most $f$~byzantine agents, receiving the same message from $f+1$~distinct agents is sufficient to guarantee the message veracity. But counting is not part of epistemic modal logic, making it hard to explain this increase in knowledge in the modal language of knowledge.

One of the byzantine distributed problems we analyzed was the Firing Rebels with Relay~\cite{Fimml17}, which is related to both the Byzantine Firing Squad by Burns and Lynch~\cite{BurnsL87} and the consistent broadcasting primitive Srikanth and Toueg~\cite{SrikanthT87JACM}, the latter used for fault-tolerant clock synchronization and byzantine synchronous consensus. In the course of this analysis~\cite{FruzsaKS21}, we discovered a new modality we called  \emph{hope}~$H_i$. In this paper, we show that $H_i\phi$~represents exactly the informational content of receiving message~$\phi$ from agent~$i$ and argue that the language with both knowledge and hope modalities for each agent is the right language for the epistemic analysis of byzantine distributed systems.

In byzantine systems, a message  is either taken at face value or ignored, depending on whether the sender is correct or byzantine.  However, many communicative situations call for a more nuanced way of interpreting information than all or nothing. For instance, the quantity maxim of Grice~\cite{Grice75} may allow to extract additional information from the message by considering what could have been said but was not. To address such more complex types of communication we propose a generalization of hope modality that we call \emph{creed}. 

The paper is structured as follows. In Sect.~\ref{sect:k}, we recall the standard representation of knowledge in distributed systems via the runs and systems framework and the resulting logic {\sf S5} for the knowledge modality. In Sect.~\ref{sect:b}, we show how our analysis of byzantine distributed systems forces to relativize knowledge to the agent's correctness, resulting in the so-called modality of belief as defeasible knowledge. In Sect.~\ref{sect:h}, we discuss our novel hope modality for representing information obtained by communication in byzantine distributed systems.  In Sect.~\ref{sect:c}, we introduce the creed modality that generalizes hope and can be used to describe communication in \emph{heterogeneous distributed systems}, with more complex agent types than just correct and byzantine. Finally, in Sect.~\ref{sect:conc}, we provide conclusions.

\section{Knowledge Modality}
\label{sect:k}

As already mentioned, Kripke models are the standard tool for semantic reasoning about modality in general and \emph{knowledge modality} in particular. Throughout the paper, we assume that there is a finite set $A = \{1,\dots,n\}$ of \emph{agents} denoted~$i, j, a, b,\dots$ and a countably infinite set~$\Prop$ of (\emph{propositional}) \emph{atoms} denoted~$p, q, p_1,\dots$
\begin{definition}[Multi-agent Kripke models]
\label{def:models}
A \emph{Kripke model} for the set~$A$ of agents is a tuple $M=\langle S, \RR, V\rangle$ that consists of a non-empty set~$S$ of \emph{(possible)} \emph{worlds}, a function $\RR \colon A \to 2^{S \times S}$ mapping each agent $i\in A$ to a binary \emph{accessibility relation} $\RR_i \subseteq S \times S$ on~$S$, and a \emph{valuation function} $V \colon \Prop \to 2^S$ mapping each atom~$p\in \Prop$ to a set~$V(p) \subseteq S$ of worlds where $p$~is true.

The truth of a formula~$\phi$ from the grammar $\phi \cce p \mid \neg \phi \mid (\phi \land \phi) \mid K_i \phi$ with~$p \in \Prop$~and $i \in A$ is recursively defined as follows: for a world~$s \in S$, we have 
$M, s \vDash p$ if{f} $s \in V(p)$, classical boolean connectives, and $M, s \vDash K_i \phi$~if{f} $M,t \vDash \phi$ whenever~$s \RR_i t$.
\end{definition}

For a given interpreted system~$I$ consisting of a set~$\R$ of runs, the corresponding Kripke model is constructed as follows:
\begin{definition}[Runs and systems]
If $\R$~is the set of runs of a given distributed system, then its \emph{global states} can be represented by pairs\/~$(r,t)$ of a run~$r \in \R$ and time instance~$t \in \mathbb{N}$, meaning that the set of possible worlds $S\ce \R \times \mathbb{N}$. It is assumed that, for each global state\/~$(r,t)$ each agent~$i \in A$ is in its \emph{local state}~$r_i(t)$. The accessibility relations\/~$\sim_i$ are defined based on local states:\/ $(r,t) \sim_i (r',t')$\, if{f}\, $r_i(t) = r'_i(t')$. The resulting Kripke model\/ $\langle\R \times \mathbb{N}, \sim, V\rangle$ describes the knowledge of agents.
\end{definition}

It is easy to see that these~$\sim_i$ are equivalence relations, hence, the logic for reasoning about knowledge in distributed systems ends up being (multimodal)~{\sf S5}. In particular, it validates factivity ${\sf t}: K_i \phi \to \phi$, positive introspection ${\sf 4}: K_i \phi \to K_i K_i \phi$, and negative introspection ${\sf 5}: \neg K_i \phi \to K_i \neg K_i \phi$ for each agent~$i \in A$. (See~\cite{Garson23} for details.)

Applying the Kripke definition of knowledge to an interpreted system $I$ yields
\begin{multline}
\label{eq:knowledge_RnS}
I, r, t \vDash K_i \phi 
\qquad \Longleftrightarrow \qquad
\Bigl(r_i(t)=r'_i(t') \quad \Longrightarrow\quad I, r', t' \vDash  \phi\Bigr).
\end{multline}
Thus, agent~$i$'s local state~$r_i(t)$ determines what $i$~knows at global state~$(r,t)$ (modulo the a priori assumptions about the whole system). This   explains why programming can be done based on agents' local states, despite it creating the erroneous impression of agent's knowledge not being taken into account, in seeming violation of the Knowledge of Preconditions Principle~(KoP). It is  this model of knowledge~\eqref{eq:knowledge_RnS} that was so successful in analyzing fault-free distributed systems~\cite{FaginHMV95}. 

As already mentioned in the introduction, this model does not easily translate to byzantine distributed systems, with the exception of certain types of so-called \emph{benign} faults, e.g.,~crashes and omissions~\cite{MosesT88,DworkM90,HalpernMW01}. This exception can also be explained via KoP. Recall that actions are determined by knowledge, which is determined by the local state.
Both crashes and (send) omissions represent the type of byzantine  behavior involving a faulty agent \emph{not} doing something assigned by the protocol, rather than \emph{doing} something in violation of the protocol. With no actions performed differently, no need arises to modify how knowledge is modeled in runs and systems, nor the agent's local state.\footnote{One could point out that inaction could be viewed as a kind of action, but  temporary inaction is how asynchrony is modeled in distributed systems, and the same methodology applies here more or less as is.} By contrast, if a faulty agent can perform a different action, it is not clear how to reflect this in its knowledge and local state. This becomes especially obvious when considering \emph{fully byzantine} agents that can perform any action. How does their knowledge determine their arbitrary actions? Does it mean that their knowledge should be inconsistent, which would be incompatible  with the logic of knowledge~{\sf S5}?

One of our first achievements~\cite{KuznetsPSF19FroCoS,KuznetsPSF19TARK} was a new framework extending the runs and systems framework with machinery for handling fully byzantine agents without losing the modeling of knowledge according to~\eqref{eq:knowledge_RnS}. Without getting into all the technical details of our solution~\cite{KuznetsPSFG19,Schloegl20}, the main idea behind it was to decouple the arbitrary actions of faulty agents from their knowledge. In our framework, all faulty actions are perpetrated by the (adversarial accept of the) environment. Thus, a faulty agent retains its knowledge according to~\eqref{eq:knowledge_RnS} but loses the control of its actions. 

Another barrier we had to overcome stemmed  from the tension between the factivity of {\sf S5}~knowledge and the necessity to represent faulty sensors and, more generally, to give faulty agents the ability to be mistaken. This was solved by shifting the knowledge corruption to the level of local states. In order to enable agents to be wrong, our framework imbues the environment with the ability to falsify events, actions, and/or messages and to record these ``fake'' events into the agent's local state. The result is that the agent's knowledge is subjectively factive, in compliance with~{\sf S5}, but may not match the objective reality, e.g.,~the agent may subjectively ``know'' that it has received a particular message even though in reality no such message  was ever sent. Of course, to ensure overall logical consistency, that requires to drop the a priori assumption that every received message must have been sent, which is routinely made in distributed systems with at most crash and omission failures.

To summarize, by extending runs and systems framework to the case of fully byzantine agents, which may have false memories, lie, and violate their protocol, our framework~\cite{KuznetsPSF19FroCoS,KuznetsPSF19TARK,KuznetsPSFG19} finally provided a model for analyzing knowledge of agents in fully byzantine distributed systems, while retaining the traditional {\sf S5} logical properties of knowledge.

Paradoxically, one of the first results of this analysis made it clear that knowledge is too strong to be used  for triggering actions in byzantine environments. For instance, a natural trigger for an agent's action would be some triggering event. However, we showed~\cite{KuznetsPSF19FroCoS} that  the subjective nature of the factivity of agents' knowledge has the following  consequence: in a fully byzantine distributed system, no agent (correct or faulty) can know that a particular event took place objectively. We proved this by formalizing the infamous \emph{brain in a vat} thought experiment~\cite{PessinG95} in our framework and showing that in any global state~$(r,t)$ each agent~$i$ has an indistinguishable global state~$(r',t')$ where $i$~is a ``brain in a vat,'' i.e.,~all actions and events recorded in $i$'s~local state~$r_i(t)=r'_i(t')$ are fake in run~$r'$: none of them took place. The inability to distinguish~$(r,t)$ from~$(r',t')$ precludes agent~$i$ from knowing in run~$r$ that any action or event took place, even if all agents are correct in run~$r$. Moreover, this result applies to a wide range of distributed systems, both asynchronous~\cite{KuznetsPSF19FroCoS} and synchronous~\cite{SchloeglSK21}.

This result might seem quite devastating in light of~KoP. How can an agent react to a triggering event if the agent cannot know that such an event happened? 

\section{Belief Modality}
\label{sect:b}

The solution is known at least since~\cite{Moses88}, where Moses proposed to define a  \emph{belief modality}  as knowledge relativized to the agent's correctness.\footnote{Albeit in a slightly different language involving \emph{indexical sets}.} In our language, this modality is represented by 
\begin{equation}
\label{eq:belief_def}
B_i \phi \ce K_i(\correct{i} \to \phi),
\end{equation} 
where $\correct{i}\in \Prop$ are finitely many designated propositional atoms, one per agent, with truth value determined by whether agent~$i$ is correct or not. Later in~\cite{MosesS93}, Moses and Shoham dubbed this modality \emph{belief as defeasible knowledge} and axiomatized it as~${\sf K45}$ with two additional axioms:
\begin{equation}
\label{eq:belief_axioms}
\correct{i} \to (B_i \phi \to \phi) 
\qquad\text{and}\qquad
B_i \correct{i},
\end{equation}
where {\sf K45}~means that $B_i$~has positive and negative introspection but not necessarily factivity. However,  the first additional axiom means that $B_i$~is factive for correct agents. Consequently, if a correct agent~$i$ performs an action based on~$B_i \phi$, then $\phi$~is guaranteed to hold. While this may not hold true for a faulty agent~$i$, those actions do not have to follow the protocol anyway. As a result, in a fully byzantine distributed system, the way to program a reaction to a triggering event~$\phi$ is to use $\correct{i}\to\phi$ as a trigger instead. While knowledge $K_i \phi$ is not achievable~\cite{KuznetsPSF19FroCoS,SchloeglSK21}, belief $B_i \phi = K_i(\correct{i} \to \phi)$ may be, and it results in  the same actions for correct agents.

The necessity to replace knowledge $K_i \phi$ with belief $B_i \phi$ in byzantine distributed systems is the first conclusion we formally derived in our framework. 

However, while belief enables one to describe agent's actions, it is not sufficient to explain how agents learn by communication. The problem lies in the ability of byzantine agents to lie: if agent~$i$ receives a message~$\phi$ from agent~$j$ and considers a possibility that this message is a lie, what can $i$~learn? Were $j$~guaranteed to be truthful, then $K_j \phi$~would be a precondition for sending~$\phi$ in fault-free systems per~KoP, from which $\phi$~would follow by factivity. As just discussed, the precondition in byzantine scenarios would be the weaker~$B_j \phi$, which is also factive for a correct~$j$. Yet, when the truthfulness assumption is removed, in particular, in the case of $j$~being fully byzantine, then even $B_j \phi$~becomes too strong. In fact, since a faulty~$j$ can send any message  independent of its knowledge, the informational content of $\phi$ would be non-existent, i.e.,~trivial. 

But communication is the main tool for solving distributed tasks. If agents cannot learn from messages, then how does communication help solve the task?

\section{Hope Modality}
\label{sect:h}

We propose the answer to this question in the form of the \emph{hope modality}~\cite{Fruzsa23}
\[
H_i \phi \quad\ce\quad \correct{i} \to B_i \phi \quad=\quad \correct{i} \to K_i (\correct{i}\to \phi).
\]
This modality emerged naturally in our analysis of the distributed task we called Firing Rebels with Relay~(FRR)~\cite{FruzsaKS21}, which is an asynchronous variant of the Byzantine Firing Squad problem~\cite{BurnsL87} and is closely related to the consistent broadcasting primitive~\cite{SrikanthT87JACM}. Only later did we understand the proper meaning of hope.

$H_j \phi$~is precisely the informational content of a message~$\phi$ received from agent~$j$ in a byzantine distributed system. Indeed, as we just discussed, two options should be considered while interpreting~$\phi$:  agent~$j$ is either correct or faulty. If $j$~is correct, then $B_j \phi$ is the precondition for sending $\phi$; if $j$~is faulty, then the precondition is trivial, i.e.,~$\top$. Collecting both possibilities, one concludes
\begin{equation}
\label{eq:hope_analysis}
(\correct{j} \to B_j \phi) \land (\neg \correct{j} \to \top) \qquad \leftrightarrow \qquad H_j \phi.
\end{equation}

The first axiomatization of hope, in the style of Moses--Shoham~\cite{MosesS93} was obtained by Fruzsa in~2019~\cite{Fruzsa23}: hope is axiomatized as ${\sf K45}$ plus the same two axioms~\eqref{eq:belief_axioms} with hope~$H_i \phi$ replacing belief~$B_i \phi$, plus one more additional axiom
\[
\neg \correct{i} \to H_i \phi.
\]

Both Moses--Shoham's axiomatization of belief and Fruzsa's initial axiomatization of hope lack the substitution property: e.g., substituting an arbitrary formula $\phi$ for $\correct{i}$ in the axiom $H_i \correct{i}$ yields $H_i \phi$. Validating all~$H_i \phi$ would have trivialized the hope modality, and would have rendered all communication  useless. Thus, the logic of hope in~\cite{Fruzsa23} is not a normal modal logic.

Fortunately, an alternative axiomatization of hope~\cite{vDitmarschFK22} can remove this obstacle. It turns out that expressing the correctness atoms  through hope as  
\begin{equation}
\label{eq:correct_def}
\correct{i} \ce \neg H_i \bot,
\end{equation} 
produces  an equivalent representation of hope that is axiomatized as~{\sf KB4}, i.e.,~a normal modal logic with axiom~{\sf 4} for positive introspection as earlier and with ${\sf b}: \phi \to H_i \neg H_i \neg \phi$ replacing~{\sf 5}.

So far we discussed four syntactic constructs to be used for describing reasoning about \emph{knowledge and communication} in byzantine distributed systems: 
\begin{compactdesc}
\item[knowledge modalities] for preconditions of actions;
\item[correctness atoms] for determining whether an agent is correct or faulty;
\item[belief modalities] for weaker preconditions of actions in byzantine settings;
\item[hope modalities] for the informational content of messages.
\end{compactdesc}
In view of~\eqref{eq:correct_def}~and~\eqref{eq:belief_def} it is clear that  knowledge and hope modalities for each agent suffice. As shown in~\cite{vDitmarschFK22}, correctness atoms cannot be expressed through knowledge. It, therefore, follows from~\eqref{eq:correct_def} that neither can hope.\looseness=-1

Thus, we posit that, at a minimum, a language suitable for reasoning about knowledge and communication in byzantine distributed systems should contain knowledge and hope modalities for each agent. 
\begin{definition}[Axiomatization of knowledge and hope~\cite{vDitmarschFK22}]
The logic\/~{\sf KH} of knowledge and hope has the following axioms and inference rules:

\centering
$\begin{array}{l@{:\quad}l@{\qquad}l@{:\quad}l}
\multicolumn{4}{c}{\text{all propositional tautologies}} \\
{\sf d}_{\rightarrow}^H & H_i \neg H_i \bot&
{\sf k}^K  & K_{i}(\phi \to \psi) \to ( K_{i}\phi \to K_{i}\psi)
\\
\multicolumn{2}{c}{}&
{\sf 4}^K  & K_{i}\phi \to K_{i}K_{i}\phi
\\
\multicolumn{2}{c}{}&
{\sf 5}^K  &\neg K_i\phi \to K_{i} \neg K_{i}\phi
\\
\multicolumn{2}{c}{}&
{\sf t}^K  & K_i\phi \to \phi
\\
{\sf MP} & \displaystyle\frac{\phi \quad \phi \to \psi}{\psi} 
&
{\sf Nec}^K & \displaystyle\frac{\phi}{K_i \phi}
\\
\multicolumn{4}{c}{{\sf kh}:  \quad H_i \phi \leftrightarrow \bigl(\neg H_i\bot \to K_i(\neg H_i\bot \to \phi)\bigr)} 
\end{array}$
\end{definition}

Kripke models for this logic are a simple generalization of models from Def.~\ref{def:models} that has two accessibility relations per agent:
\begin{definition}[Kripke models for knowledge and hope~\cite{vDitmarschFK22}]
\label{def:logic_kno_hope}
A \emph{Kripke model for knowledge and hope} for the set~$A$ of agents is a tuple $M = \langle S, \KK, \HH, V\rangle$ that consists of a set~$S\ne \varnothing$, a valuation function $V \colon \Prop \to 2^S$, and functions  $\KK,\HH \colon A \to 2^{S \times S}$ mapping each agent~$i\in A$ to  binary accessibility relations~$\KK_i$~and~$\HH_i$  on~$S$ respectively such that each~$\KK_i$ is an equivalence relation (reflexive, transitive, and symmetric), each~$\HH_i$ is a partial equivalence relation (transitive and symmetric), and, in addition, $\HH_i \subseteq \KK_i$ for each~$i \in A$ and $s \KK_i t$ implies $s \HH_i t$ whenever $s \HH_i s'$ and $t \HH_i t'$ for some worlds~$s'$~and~$t'$ for each~$i\in A$.

The truth of a formula~$\phi$ from the grammar $\phi \cce p \mid \neg \phi \mid (\phi \land \phi) \mid K_i \phi \mid H_i \phi$ with~$p \in \Prop$~and $i \in A$ is the same as in Def.~\ref{def:models} except $M, s \vDash K_i \phi$~if{f} $M,t \vDash \phi$ whenever~$s \KK_i t$, with the new clause $M, s \vDash H_i \phi$~if{f} $M,t \vDash \phi$ whenever~$s \HH_i t$.
\end{definition}

We illustrate how to model communication in byzantine distributed systems:\looseness=-1 
\begin{example}
It is standard in distributed systems to restrict the number of byzantine agents to  at most~$f$ for some $0 \leq f \leq n$. Indeed, such a restriction is typically necessary for the distributed task at hand to be solvable, with $n\geq 2f+1$ or $n\geq 3f+1$ being a common necessary condition.
In light of~\eqref{eq:correct_def}, this can be formalized as
\[
\Byzf \ce \bigvee\limits_{G\subseteq A \atop |G|=n-f}\,\,\bigwedge\limits_{i \in G}\neg H_i\bot.
\] 
Using the standard group notion of mutual hope $E^H_G \phi \ce \bigwedge_{i \in G} H_i \phi$  for a group~$G \subseteq A$ of agents, we can derive factivity of mutual hope for sufficiently large groups: whenever $|G|\geq f+1$, 
\begin{equation}
\label{eq:byz_learning}
{\sf KH} + \Byzf \quad\vdash\quad E^H_G \phi \to \phi.
\end{equation}
As simple as the derivation of~\eqref{eq:byz_learning} might seem, it faithfully formalizes the process of learning in byzantine distributed systems: if the same message is received from $f+1$~distinct agents, then one of them must be correct, ensuring the veracity of the message.
\end{example}
Other examples formalizing specifications of byzantine systems  by knowledge and hope, as well as other properties derived using the logic from Def.~\ref{def:logic_kno_hope} can be found in~\cite{vDitmarschFK22}.

More syntactic constructs may be added to the logic of knowledge and hope as needed. For instance, our analysis of the FRR problem~\cite{FruzsaKS21} is done in a language with the addition of temporal modality $\Diamond \phi$ for ``$\phi$ holds at some point in the future'' and a mixed temporal--epistemic fixpoint group modality $C^{\Diamond H} \phi$ of \emph{eventual common hope}. Another example can be found in~\cite{vDitmarschFKS24IJCAR}, where dynamic operators in the style of Dynamic Epistemic Logic~\cite{vDitmarschvdHK07} are used to (self-)correct agents.\looseness=-1

One last thing to discuss about hope is the origins of the name. Initially, hope was chosen as an attitude similar to but weaker than belief. Indeed, in our logic, ${\sf KH} \vdash B_i \phi \to H_i \phi$. However, it is reasonable to ask whether one would actually say ``hope'' in colloquial speech in the same situations where our hope modality is employed. Here is one such example. It often happens that, meeting someone familiar after a long break, I may not be sure about the name of their child. Say, the name ``Florian'' comes to mind, but it might also be ``Miro.'' I clearly do not \emph{know} it is Florian. But, in fact, my memory may be so fuzzy that I cannot even say that I \emph{believe} it to be Florian: it may be more that, in my mind, it is, say, Florian with 85\% probability. Sometimes, in such situations, I still venture, in the name of politeness, to ask about Florian, \emph{hoping} that I got the name right. And that is the meaning of hope that our hope modality embodies.

\section{Creed Modality}
\label{sect:c}
The breakdown of hope from~\eqref{eq:hope_analysis} underscores the all-or-nothing dichotomy of correct/faulty agents: correct agents are completely trusted,  while faulty agents are completely distrusted. The same approach, however, can also be applied in less discriminating circumstances. In this section, we show how hope can be generalized from byzantine distributed systems, where agents are either  correct or faulty, to \emph{heterogeneous distributed systems} where agents may belong to several different types with each type communicating in its own way that need not be fully known to other agent types. We start by illustrating the utility of this approach  by solving  one of Smullyan's puzzles~\cite[Puzzle~28]{Smullyan78}.

\begin{example}[Knights and knaves~\cite{CignaraleKRS23}]
 All inhabitants of an island are either
\emph{knights} who always tell the truth or \emph{knaves} who always lie. Agents~$a$~and~$b$ are two of these inhabitants. Agent~$a$ makes the following statement: ``At least one of us two is a knave.''
What are~$a$~and~$b$?
 
Here we have two types of agents: knights and knaves.
Thinking of utterances as actions that have preconditions, we can reformulate the definition of these two types as follows: the precondition for agent~$j$ saying~$\phi$ is \looseness=-1
\begin{compactitem}
\item $\phi$, and, according to KoP, $K_j \phi$~for a knight;
\item $\neg \phi$, and, according to KoP, $K_j \neg \phi$~for a knave.
\end{compactitem}
Consequently, the analog of~\eqref{eq:hope_analysis} for agent~$j$'s utterance of~$\phi$ would in this case be
\[
(\kni_j \to K_j\phi) \land (\kna_j \to K_j \neg \phi),
\]
where $\kni_j$ and $\kna_j$ are atoms that, like $\correct{j}$ for hope, signify the type of agent~$j$. The above statement of agent~$a$ can be formalized in this language as $\phi \ce  \kna_a \lor \kna_b$. Accordingly, the informational content of this $\phi$ amounts to 
\begin{equation}
\label{eq:kni_kna}
\bigl(\kni_a \to K_a(\kna_a \lor \kna_b)\bigr) \land \bigl(\kna_a \to K_a\neg (\kna_a \lor \kna_b)\bigr).
\end{equation}
Using normal modal reasoning, factivity of knowledge, and the puzzle's a priori assumption $\kni_j \leftrightarrow \neg \kna_j$, we can easily  derive $\kni_a \land \kna_b$ from~\eqref{eq:kni_kna}. This provides the answer to the puzzle: $a$~is a knight while $b$~is a knave.
\end{example} 

In this vein, we defined \emph{creed modality}~\cite{CignaraleKRS23} as a generalization of hope:
\[
\creed{a}{L}{S} \phi \ce S_a \to K_a f_{LS}(\phi)
\]
where $L$~is the type of the listening agent, $S$~is a possible type the speaking agent~$a$ might have, and $f_{LS}(\phi)$ is the strongest precondition for speaking $\phi$ by an $S$-type agent that $L$-type agents are aware of. Using creed, an $L$-type agent can extract the following informational content from a message $\phi$ of agent $a$:
\[
\creed{a}{L}{S_1} \phi \land \dots \land \creed{a}{L}{S_k} \phi
\]
where $S_1, \dots, S_k$ are all the agent types the listener thinks $a$ may belong to.

\section{Conclusions}
\label{sect:conc}

In this paper, we have proposed the minimal language needed to reason about knowledge and its increase via communication in byzantine distributed systems. This language should include, for each agent, knowledge modality to describe preconditions for this agent's actions and hope modality to describe what other agents learn from this agent's messages. Hope modality first appeared in the epistemic analysis of a particular distributed task, a simplified version of consistent broadcasting, further proving its indispensability for reasoning about knowledge and communication in distributed systems with fully byzantine agents.

Looking beyond hope, we  have shown the first glimpse of how to logically model communication among agents of different types and different communication strategies, the situation that is to become more and more widespread as distributed systems grow both in size and in complexity. The research into this novel creed modality is only in its initial stages.

\begin{credits}
\subsubsection{\ackname} 
This paper summarizes a particular strain of the results obtained within the framework of the Austrian Science Fund (FWF) project ByzDEL (P33600). Many people contributed to our research project and shared their ideas and expertise over the years, whether officially or unofficially participating in or simply collaborating with the project. Some are cited. Others collaborated on other topics  or contributed in less visible but nevertheless tangible and crucial ways. I will provide an alphabetical list of people involved in the project in all these various capacities, to whom I am immensely thankful for joining me on this exciting ride:
Giorgio Cignarale, Hans van Ditmarsch, Stephan Felber, Patrik Fimml, Krisztina Fruzsa, Lucas Gr\'eaux, Yoram Moses, Laurent Prosperi, Sergio Rajsbaum, Rojo Randrianomentsoa, Hugo Rinc\'on Galeana, Thomas Schl\"ogl, Ulrich Schmid, Thomas Studer, and Tuomas Tahko.\looseness=-1
\end{credits}

\bibliographystyle{abbrvurl}
\newcommand{\BenthemSortNoop}[1]{}\newcommand{\DitmarschSortNoop}[1]{}\newcommand{\EijckSortNoop}[1]{}\newcommand{\HoekSortNoop}[1]{}\newcommand{\MeydenSortNoop}[1]{}

\end{document}